\documentclass[12pt,preprint]{aastex}

\shorttitle{A-X infrared bands of AlO: new detections}
\shortauthors{Banerjee et al.}

\begin{document}

\title{The A-X infrared bands of Aluminum Oxide  in stars: search and new detections.}

\author{D. P. K. Banerjee$^1$, W. P. Varricatt$^2$, Blesson Mathew$^1$, O. Launila$^3$, N. M. Ashok$^1$}

\affil{$^1$Astronomy and Astrophysics Division, Physical Research Laboratory,
Navrangpura, Ahmedabad  - 380 009, Gujarat, India}
\email{orion@prl.res.in}
\affil{$^2$Joint Astronomy Centre, 660 N. Aohoku Place, University Park, Hilo,  Hawaii - 96720,  USA}

\affil{$^3$KTH-AlbaNova, Applied Physics, Roslagstullsbacken 21, 106 91 Stockholm, Sweden}

\begin{abstract}
We describe a search for the A-X infrared bands of AlO with a view to better understand
the characteristics of this radical.
These bands are infrequently encountered in astronomical sources but surprisingly were very
prominent in the spectra of two well-known, nova-like variables (V838 Mon and V4332 Sgr) thereby
motivating us to explore the physical conditions necessary for their excitation.
In this study, we present the detection of A-X bands in the spectra of
13 out of 17 stars, selected on the basis of their J-K colors as potential
candidates for detection of these bands.
The majority of  the AlO detections are in AGB stars viz. 9 OH/IR stars, 2 Mira variables and 2
 bright infrared sources.
Our study shows that the A-X bands are fairly prevalent in sources with
low temperature and O-rich environments.
Interesting variation in strength of the AlO bands in one of the sources (IRAS 18530+0817)
is reported and the cause for this is examined.
Possible applications of the present study are discussed in terms of the role of
AlO in alumina dust formation, the scope for estimating the radioactive $^{26}$Al
content in AGB stars from the A-X bands, and providing possible targets for
further mm/radio studies of AlO which has recently been discovered at millimeter wavelengths.
\end{abstract}

\keywords{Infrared: stars --- Stars: AGB and post-AGB --- ISM: molecules --- Techniques: spectroscopic}

\section{Introduction}
In this study we describe a search to detect the infrared A-X bands of the AlO
radical. While several oxide molecules like  TiO and VO are routinely encountered
in the spectra of cool O-rich  stars, AlO is detected rarely and is hence a
poorly studied molecule. But, as we proceed to show,  AlO could  be detected
more often if searched for in appropriate sources.
The motivation for the study arose from two eruptive variables V838 Mon and V4332 Sgr which have been
objects of great interest in recent years. Both objects underwent initial large-amplitude
nova-like outbursts  but it was soon realized that they were
different from classical novae or other known classes of CVs (e.g. Munari et
al. 2002). Worth mentioning specifically is the haloed status achieved by
V838 Mon by virtue of the striking light echo that adorned it \citep{Bond03,Banerjee06}.
A general consensus exists today that they belong to a new class of eruptive
variables called intermediate luminosity red transients (ILRTs; \citet{Bond11}).
With the discovery of several more such red transients in recent times
\citep{Humphreys11}, and the likelihood of more discoveries emerging
from ongoing synoptic sky surveys, there is currently a great deal of interest in such objects.

Our concern here is with the unusual presence of rarely seen molecular bands of the A-X system of
AlO radical, which are strongly seen in the near-infrared spectra of V838 Mon
and V4332 Sgr and which could potentially be expected in other  ILRTs too.
In V838 Mon, these AlO bands were seen in absorption \citep{Evans03,Lynch04,Banerjee05};
in V4332 Sgr they were spectacularly seen in emission \citep{Banerjee03}.
Barring V838 Mon, V4332 Sgr and IRAS 18530+0817 \citep{Walker97,Evans03},
a search in the literature shows a severe
paucity of A-X band detections of AlO in other sources and
a strong need therefore arose to understand why the A-X bands were detected so infrequently.
It was perceived that a statistically larger number of detections of the band
would help to understand  the characteristics and properties of AlO better.

An AlO detection could also help in exploring its role in the formation of
alumina dust. Alumina is considered to be the earliest dust
condensate in O-rich stars - condensing out at
1760K. In the chaotic silicate hypothesis, the AlO radical plays a pivotal role
in silicate and alumina dust formation (Nuth $\&$ Hecht,
1990). Observationally too, it was intriguing to see
strong AlO emission in V4332 Sgr being simultaneously accompanied by a strong 11 $\mu$m alumina
feature \citep{Banerjee07}. But the interconnection between
a gas phase component (AlO) and a solid state component (alumina) needs
further study and a statistically large sample of AlO detections should be
helpful. From the observational point of view, it is timely to plan or carry
out such observations  with the present availability of IR space-borne
facilities like SOFIA (Stratospheric Observatory for
Infrared Astronomy)\footnote{www.sofia.usra.edu} and of the
James Webb Space Telescope\footnote{www.jwst.nasa.gov} in the near future.
While the mid-IR dust features could be studied by the space-borne facilities,
the near-IR AlO bands could be simultaneously studied from the ground.
The present study could also help explore an important issue viz. the
radioactive $^{26}$Al content in AGB stars as discussed in section 3.
Finally, it should also be noted that the first millimeter detection of AlO
has been recently made in the O-rich supergiant star VY CMa
\citep{Tenenbaum09} which could lead to similar searches in other objects. The sources detected
here could hence be potential targets for such further mm/radio detections;
the different transitions occurring in the mm/radio regimes are given in \citet{Launila09}.

It may be noted that a search for near-IR AlO bands had been made earlier
too. \citet{Luck74} had looked for the (1,0) bands in selected supergiants and Mira variables
but failed to detect them due to miscalculated wavelength positions of the
bands (the molecular constants were poorly known then). The wavelengths of the
different A-X bands are now well determined from the detailed study of the AlO radical by \citet{Launila94}.

The principal result reported here is the near-IR detection of several bands
of AlO in the 1.0--1.35 micron region. These detections are made in 13 out of
17 pre-selected sources thereby  giving a high detection  rate of
almost 75 percent.  Our study shows that the A-X bands are thus not so rare
and that they should be fairly prevalent in low temperature, O-rich environments.

\section{Observations \& Analysis}
The strategy for source selection was based on the
excitation conditions for the A-X bands which arise from transitions between different vibrational
levels of the excited A and the ground X electronic
states. The vibrational states are low-lying, typically 0.6 to 1 eV above ground level
\citep{Launila94} and are thus  expected to be favorably excited at low
temperatures. This is consistent with their appearance in V838 Mon and V4332 Sgr, both of which have
cool dusty envelopes with T $<$ 1000K \citep{Lynch04,Banerjee03}.
AlO emission may therefore be expected from similar stellar environments viz. the dusty circumstellar
envelopes of evolved stars. A criterion that the $(J-K)$
color should be large was hence applied to ensure selection of such sources.
We set $J-K$ $>$ 4 and chose bright sources to optimize telescope time
requirement by setting $J$ $<$ 10. A search of the
2MASS point source catalog with these constraints yielded 487 sources.
The sample was reduced by ensuring the sources showed the 10 and
18 $\mu$m silicate features in their IRAS spectra (Olnon et
al. 1986) indicating the O-rich environment necessary for AlO. A further
constraint was that the 10 $\mu$m  silicate feature be broad or display a clearly discernible 11 $\mu$m
alumina feature on its red wing. Such profiles, it has been argued
\citep{Speck00,Banerjee07}, suggest the presence of alumina and hence
strengthen the possibility of AlO being present. A final reduction to 17
sources was made considering source accessibility (RA and $\delta$
constraints; $\delta$  was set $>$ $-$40 degrees) from
the 3.8m UK Infrared Telescope (UKIRT) where the observations were planned and made.

Observations were performed with the UKIRT 1-5 micron Imager Spectrometer (UIST),
using an IJ grism and a 4-pix (0.48")-wide slit; the spectra obtained cover a
wavelength range of 0.86 $\mu$m -- 1.4 $\mu$m with a spectral resolution of $\sim$320.
Observations of a flat, arc and an early type (B9V-A2V) ratioing star
were obtained with the same instrument setting as for the program objects
before each set of observations.
Before rationing the spectrum of the source with the spectrum of the
telluric standard, the hydrogen Pa $\beta$ and Pa $\gamma$ absorption features
in the spectrum of the latter were removed. The initial processing of the
data were carried out using the UKIRT pipeline
ORACDR; the extraction and calibration of the spectra were carried out using
STARLINK-FIGARO and IRAF.  The log of the observations
with some source details is given in Table 1.

\begin{figure}
\includegraphics[bb=30 164 270 690, width=3.5in,height=8.15in, clip]{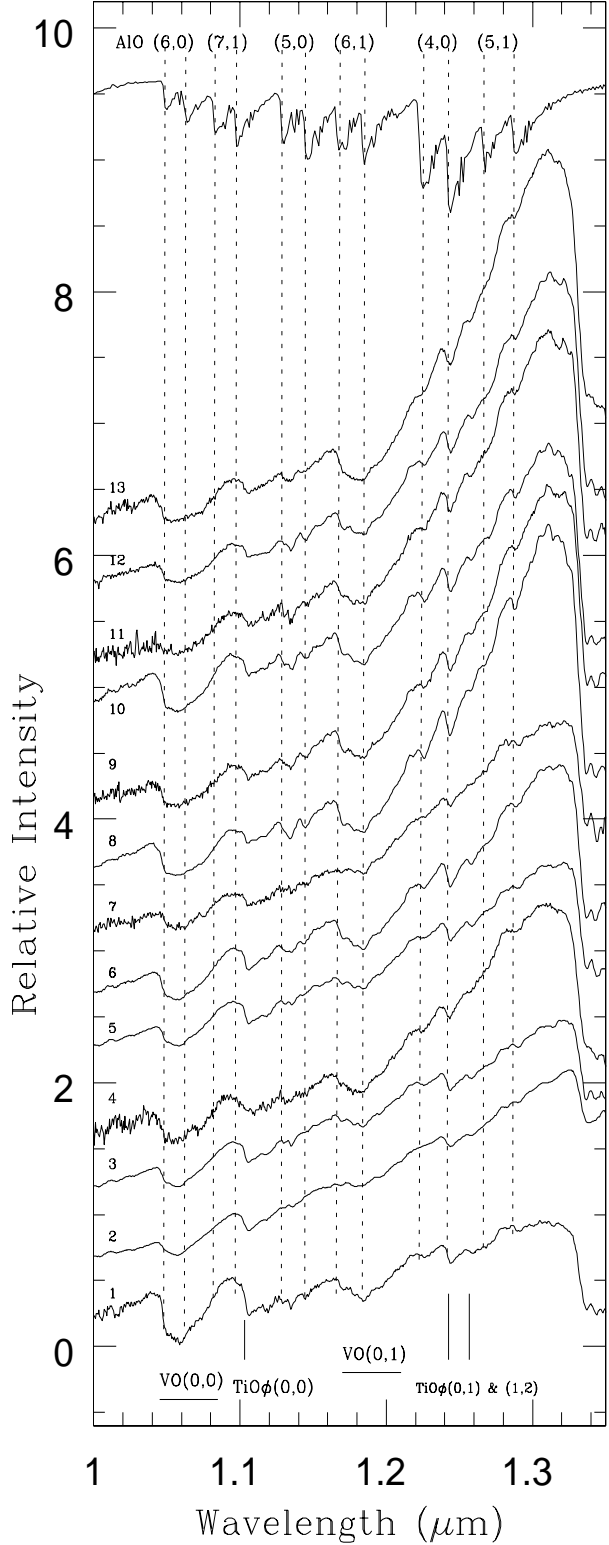}
\caption{\scriptsize Spectra in the  1--1.35 $\mu$m region of 13 stars which show AlO features.
A synthetic spectrum of AlO is displayed at top with drop lines and labels showing the origin
of the different A-X bands. The position of the relevant features of VO and TiO are
shown at the bottom. The label numbers of the spectra and the associated source IRAS
designations are: 1 - 19178-2620; 2 - 20217+3330;
3 - 18373-0021; 4 - 20267+2105; 5 - 18476+0555; 6 - 18481-0346;
7 - 19043+1009; 8 - 05073+5248; 9 - 18027-2314; 10 - 18069+0911;
11 - 18091-1656 12 -  18120-1417; 13 - 20077-0625. }
\end{figure}

\begin{figure}
\includegraphics[scale=0.5]{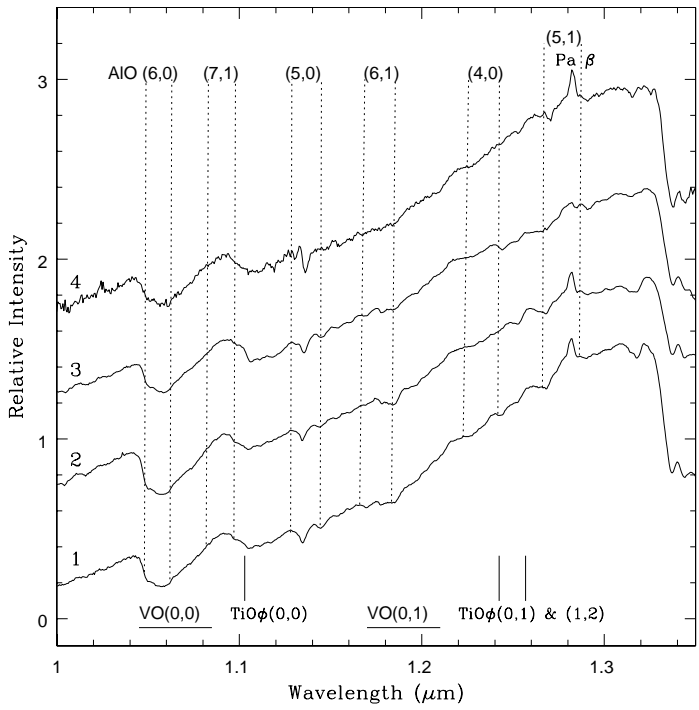}
\caption{Spectra of IRAS 17194-3354, 17209-3126, 18139-1811 and
18530+0817 (labelled 1, 2, 3 and 4 respectively) where AlO features were not detected clearly.
As in Figure 1, the expected positions of the relevant molecular features are shown.}
\end{figure}

\begin{table}
\begin{center}
\caption{Journal of observations \label{tbl-1}}
\begin{tabular}{ccccccc}
\tableline\tableline
IRAS desi- & Obs. & Exp. & 2MASS & Airmass \\
gnation\tablenotemark{1} & Date & (s) &  ($J,H,K$) & Source Std. \\
\tableline
05073+5248$^g$ & 11-08-2008 & 480 & 8.204, 5.775, 3.861 & 1.496 1.543\\
17194-3354$^d$ & 29-07-2008 & 480 & 9.955, 7.332, 5.671 & 1.730 1.381\\
17209-3126$^d$ & 29-07-2008 & 480 & 9.847, 7.287, 5.532 & 1.659 1.381\\	
18027-2314$^d$ & 29-07-2008 & 720 & 9.474, 6.825, 5.158 & 1.585 1.381\\
18069+0911$^a$ & 25-06-2008 & 120 & 9.449, 6.900, 5.137 & 1.039 1.084\\
18091-1656$^e$ & 28-06-2008 & 400 & 9.204, 6.613, 5.057 & 1.296 1.090\\	
18120-1417$^d$ & 29-07-2008 & 540 & 8.552, 6.004, 4.523 & 1.477 1.381\\
18139-1811$^d$ & 29-07-2008 & 480 & 9.673, 7.067, 5.372 & 1.634 1.381\\
18373-0021$^e$ & 28-06-2008 & 160 & 9.446, 6.498, 4.595 & 1.112 1.031\\
18476+0555$^e$ & 28-06-2008 & 480 & 9.424, 6.958, 5.245 & 1.047 1.031\\
18481-0346$^e$ & 28-06-2008 & 600 & 9.775, 7.271, 5.571 & 1.132 1.031\\
18530+0817$^a$ & 25-06-2008 & 160 & 9.404, 6.905, 5.119 & 1.143 1.084\\
19043+1009$^c$ & 25-06-2008 & 160 & 9.355, 6.990, 5.259 & 1.071 1.092\\
19178-2620$^b$ & 25-06-2008 & 160 & 9.643, 6.720, 4.588 & 1.639 1.200\\
20077-0625$^f$ & 28-06-2008 & 120 & 6.906, 3.923, 2.059 & 1.124 1.008\\
20217+3330$^f$ & 28-06-2008 & 540 & 8.768, 6.239, 4.674 & 1.031 1.008\\
20267+2105$^f$ & 28-06-2008 & 540 & 8.792, 6.310, 4.340 & 1.001 1.008\\
\tableline
\end{tabular}
\tablenotetext{1}{Telluric standards used: a - BS 6977 (A0Vn), b - BS 7400
(A0V), c - BS 7390 (A0V), d - BS 6581 (A2V), e - BS 6992 (B9V),
f - BS 7891 (A0V), g - BS 1692 (A0V).
Mean airmass of the source and standard are given in the last 2 columns.}

\end{center}
\end{table}

\begin{deluxetable}{crrrrrrr}
\tabletypesize{\scriptsize}
\tablewidth{0pt}
\tablecaption{Source descriptions\tablenotemark{1} \label{tbl-2}}
\tablehead{
\colhead{IRAS desi-} & \multicolumn{3}{c}{SiO}  & \multicolumn{2}{c}{OH}  &
\colhead{H$_2$O maser} & \colhead{Source} \\
\colhead{gnation}   & \colhead{v=1, J=1-0} & \colhead{v=2, J=1-0} &
\colhead{v=1, J=2-1}  & \colhead{1612}  & \colhead{1667}  & \colhead{}  & \colhead{details}\\
\colhead{}  & \colhead{43.12 GHz} & \colhead{42.82 GHz} & \colhead{86.24 GHz}
& \colhead{MHz} & \colhead{MHz} & \colhead{22.2 GHz} & \colhead{}
}
\startdata
\label{sourcedetails}
05073+5248 & $\checkmark$(1) & $\checkmark$(1)  &    & $\checkmark$(2) &
$\checkmark$(3) &     $\checkmark$(1) & Mira, Sp. class M10, $P$ = 635d (4) \\
17194-3354 & &   &    &  &  &    & IR source, poorly studied (5) \\
17209-3126 & & & & $\times$(2) & & & IR source, poorly studied (5) \\
18027-2314 & $\checkmark$(6) & $\checkmark$(6) & & $\checkmark$(2) & & & OH/IR
star (2)\tablenotemark{2}\\
18069+0911 &  $\checkmark$(1) & $\checkmark$(1) & & & & $\checkmark$(1) &
Mira, $P$ unknown (1,4) \\
18091-1656 &  $\checkmark$(6) & $\checkmark$(6) & &  $\checkmark$(17) & & &
OH/IR star (17){\bf \tablenotemark{3}}\\
18120-1417 & $\times$(8) & $\times$(8) & & $\checkmark$(17)
& $\checkmark$(3) & & OH/IR star (3,17)\\
18139-1811 & & & & $\checkmark$(2) & $\times$(3) & & OH/IR star (2,5)\\
18373-0021 & &   &    &  &  &    & IR source, poorly studied (5) \\
18476+0555 & &   &    & $\checkmark$(7) &  &    & OH/IR star (7)\\
18481-0346 & $\checkmark$(8) & $\checkmark$(8)  &    &  &  &    & IR source (5) \\
18530+0817 & & & & $\times$(9) & & & Mira? (10)\tablenotemark{4} \\
19043+1009 & & & &  $\checkmark$(17) & & $\checkmark$(11) & OH/IR star (17)\\
19178-2620 & $\checkmark$(13) & $\checkmark$(13) & $\checkmark$(13) &
 $\checkmark$(17) & $\checkmark$(3) & $\times$(12) & OH/IR star (3,17)\\
20077-0625 & $\checkmark$(14) & $\checkmark$(14) & $\checkmark$(14,15) &
$\checkmark$(2) & & & OH/IR star (2)\tablenotemark{5}\\
20217+3330 & & & & $\times$(9) & & & IR source (5)\\
20267+2105 & & & $\checkmark$(18) & & $\checkmark$(3) & & OH/IR star (3)\\
\enddata
\tablenotetext{1}{A check mark indicates a detection and a cross an
  attempted search followed by a non-detection. The numbers in parantheses are
  the relevant references which are (1) \citet{Kim10} (2)
  \citet{teLintelHekkert91} (3) \citet{LeSqueren92} (4) \citet{Benson90} (5)
  SIMBAD database (6) \citet{Deguchi00} (7) \citet{Eder88} (8)
  \citet{Deguchi04} (9) \citet{Lewis90} (10) \citet{Walker97} (11)
  \citet{Cesaroni88} (12) \citet{Deguchi89} (13) \citet{Nyman98} (14)
  \citet{Spencer81} (15) \citet{LeBertre90} (16) \citet{Hansen75}
 (17) \citet{teLintelHekkert89} (18) \citet{Cernicharo97}} 
\tablenotetext{2}{The $JHK$ magnitudes are found to be variable as expected for OH/IR stars,
between 2MASS values and $J$ = 13.784, $H$ = 8.444, $K$ = 6.241
as reported by \citet{Deguchi02}.}
\tablenotetext{3}{The $JHK$ magnitudes are found to vary between 2MASS values
and $H$ = 6.654, $K$ = 5.631 as reported  by \citet{Deguchi02}.}
\tablenotetext{4}{A detailed infrared study has been done by \citet{Walker97}
who suggest it could be Mira variable; discussed further in the text.}
\tablenotetext{5}{ \citet{LeBertre90} give a period of $\sim$ 650 days,
SIMBAD lists it as a Mira with $V$ = 20 from the study of \citet{Hansen75}.}
\end{deluxetable}

\begin{table}
\begin{center}
\caption{A$^2\Pi_i$ - X$^2\Sigma^+$ bandheads of AlO
in the 1 to 1.35 $\mu$m region\label{tbl-3}}
\begin{tabular}{@{}lrr@{}}
\tableline
Band &  Wavelength ($\mu$m) A$^2\Pi_{3/2}$  &	Wavelength ($\mu$m)
A$^2\Pi_{1/2}$ \\
\tableline
 (3,0)  & 1.3589   & 1.3358 \\
 (4,0)  & 1.2417   & 1.2226 \\
 (5,0)  & 1.1443   & 1.1281 \\
 (5,1)  & 1.2863   & 1.2661 \\
 (6,0)  & 1.0619    & 1.0480  \\
 (6,1)  & 1.1834    & 1.1659  \\
 (7,1)  & 1.0969    & 1.0818  \\
\tableline
\end{tabular}
\end{center}
\end{table}

\section{Results}
Before presenting the spectra proper, a brief description of the sources is given below
to show the nature of the selected sources. The  sources are in general
heavily obscured with  very faint or no optical
counterpart in the Digital Sky Survey plates. Most of them show maser emission
in one or more of the relevant masing lines of SiO, OH or H$_2$O.
The lines in which such detections have been made and the classification
details of the source are summarized in Table 2 along with a few more details.

Figure 1 shows  the spectra of 13 stars where AlO is clearly detected while
Figure 2 shows the 4 non-detections. Of the 13 stars with detections,
2 are Mira variables, 9 are OH/IR stars and 2 are classified as infrared sources
in SIMBAD. Of the 4 non-detections two are infrared sources, one is an OH/IR star and
one is possibly a Mira variable. At the top of the Figure 1, an inverted synthetic emission spectrum
of AlO is shown whose primary purpose is to show the expected positions of the AlO bands.
The wavelengths of the major bandheads of the A-X system in the 1--1.35
micron region are listed in Table 3. Drop lines at the position of the band
origins are also shown. The clearest detections of AlO are in the (4,0) bands at 1.2417
and 1.2226 $\mu$m. The 1.2863 $\mu$m component of the (5,1) band is also clearly seen
while a small but discernible dip is also seen at the position of the counterpart 1.2661 $\mu$m feature.
Also there are discernible bandheads at the predicted positions of the (5,0) bands at 1.1443 and 1.1281 $\mu$m
indicating that these bands are also present. We have compared our spectra with those
of \citet{Joyce98} and \citet{Hinkle89} wherein a comprehensive identification of the
$J$ band molecular features in cool stars is given. For the TiO bands  we have
also referred to \citet{Jorgensen94} and \citet{Galehouse80}.
From these studies, it is confirmed that no other molecular
features are expected to be present at the positions of the (4,0), (5,0) and (5,1) bands detected
here, and discussed above,  thereby strongly suggesting they are due to AlO.
The only case of overlapping of features
is between  the 1.2424 $\mu$m  component of the TiO $\phi$ (0,1) system  and
the 1.2417 $\mu$m AlO feature. Using the molecular constants from Jorgensen (1994; see figure 5),
the $\phi$ system is expected to produce the (0,1) and (1,2) band heads at 1.2424 and
1.2568 $\mu$m respectively. In the spectra, the 1.2568 $\mu$m TiO bandhead
is weak but is still consistently  seen in the spectra (its position is marked in Figure 1).
The only other TiO feature that is seen is the $\phi$ (0,0) system with its bandhead at 1.1035 $\mu$m.

The deep absorption  at the red end of the $J$ band due to water in the source makes
it difficult to comment on the possible presence of the (3,0) bands at 1.3589 and 1.3358 $\mu$m.
The 1.3358 $\mu$m band was seen in V4332 Sgr where water absorption was weak \citep{Banerjee03}.
VO has two prominent A-X bands in this region namely VO (0,0) with its bandhead at 1.056 $\mu$m
and VO(1,1) extending between 1.1682 $\mu$m to 1.2158 $\mu$m \citep{Cheung82}.
Both these bands are seen in the spectra of several stars \citep{Joyce98,Hinkle89}.
We suspect that the VO(0,1) band is present but strongly contaminated with the AlO (6,1) bands.
The reason for this is that VO (0,1)  is known not to have any conspicuous bandhead
as pointed out by Hinkle et al. (1989, see Table 2 therein) who therefore while assigning a
wavelength to this band give a central frequency rather than a bandhead frequency.
The spectra of R Cas in Joyce et al. (1998, Figure 3 therein) also supports this where
it is seen that while VO (0,0) has a sharp bandhead, VO (0,1) on the other hand
displays a broad feature without a significant bandhead. However in our spectra
a prominent bandhead is seen at the onset of the VO (0,1) band which we feel is instead
due to the 1.1659 $\mu$m band of AlO (6,1). In addition, there is also a clear indentation
in the VO (0,1) feature at $\sim$ 1.184 $\mu$m that coincides with the 1.1834 AlO (6,1)
component. We hence infer that although
VO (0,1) may be present, the (6,1) bands of AlO are also present in the 1.16 to 1.21 $\mu$m region.

In many of the spectra in Figures 1 and 2, especially those in the
latter, an absorption feature at  1.1345 $\mu$m  appears to be seen which does not
correspond with any of the expected molecular features. This is likely to be
an artifact arising from a strong H$_2$O telluric absorption feature expected
at this position. This feature appears to have been incompletely removed in
the ratioing process due to the slight airmass differences between the source
and standard star during observations (as indicated in Table 1).

An intriguing result is the variation seen in the spectrum of the source
IRAS 18530+0817. This is the only source from our list which has been studied in detail
in the near and mid-infrared earlier \citep{Walker97}. A comparison between our $J$ band spectrum
and a similar one taken on 19 May 1995 by  Walker et al. (1997, Figure 2a therein)   shows considerable
changes. AlO absorption bands are prominently seen in the 1995 spectrum but
have disappeared in our spectrum (the source is included in the non-detections in Figure 2).
The H Paschen $\beta$ line at 1.2818 $\mu$m is also seen in emission in our data.
Interestingly the 10 $\mu$m silicate profile, which shows a clear alumina component at 11 $\mu$m,
is also found to vary. Most striking is the change in the 10 $\mu$m profile
in the two IRAS spectra, taken in 1983 and separated by 6 months,  wherein a
large decrease in intensity level as well as striking
changes in the profile morphology are seen. The bright  emission feature at
9.5 $\mu$m (due to silicates) remains
relatively unaltered whereas the emission component around 11$\mu$m
(presumably due to alumina) has almost disappeared. Since AlO is also
established to be varying in the source, it will be interesting to take
simultaneous $IJ$ spectra of the AlO bands as well as that of the 11 $\mu$m
alumina feature and see what kind of a correlation is found. From an
astrochemistry point of view, this may be a meaningful study as discussed in section 1.

\section{Discussion}
The variation of the AlO band strengths in IRAS 18530+0817 may be a sign of a
more generic trait in variability. It may be  related to the anomalous
variation of the optical B-X bands of AlO, notably the (0,0) 4842 \AA~band, in
Mira variables reported by \citet{Keenan69}. \citet{Keenan69} conclude that
the strength of the AlO bands can be quite different in two Mira variables of
the same temperature and the same star observed at two different maxima can
have extremely different AlO strengths (hence the anomalous behavior).
This variation in strengths from cycle to cycle is not shared by other
molecules like ScO and VO which vary predictably with spectral type.
The bands are occasionally found to go into emission from absorption and we
speculate that this must be similarly happening to IRAS 18530+0817, and
possibly the other stars with non-detections in Figure 2, wherein the AlO
absorption features may have been filled in by emission (the Paschen $\beta$
line going into emission could be an indication of this).
The anomalous behavior of the B-X bands is still an unexplained phenomenon.

An AlO detection could open the door for a subsequent search for radioactive $^{26}$Al in the object.
The same A-X band of AlO, with the Al atom therein being
either $^{26}$Al or $^{27}$Al, can be sufficiently separated in wavelength to be resolved at
intermediate/high spectral resolution \citep{Banerjee04}. For example, our calculations show that the
origin of the (4,0) bands of $^{27}$Al$^{16}$O  and $^{26}$Al$^{16}$O occur at 1.22558 and 1.22053 $\mu$m
respectively and hence resolving even individual rotational lines should be possible at higher resolution.
The yield of $^{26}$Al/$^{27}$Al in AGB stars has been the subject of
considerable study (e.g reviews by \citet{Prantzos96}, \citet{Lugaro08}).
It may be noted that the bulk of our positive AlO detections are in OH/IR stars and Mira variables which are
essentially AGB stars. The OH/IR stars in particular are very evolved stars
belonging to the tip of the AGB branch enroute to their becoming planetary
nebulae.  Low mass AGB stars ($\le$ 3 to 4 M$_\odot$)
produce relatively smaller amounts of $^{26}$Al primarily from hydrogen shell
burning \citep{Lugaro08} while $^{26}$Al production is enhanced in high mass
and super AGB stars \citep{Siess08} due to the process of hot bottom burning
and the third dredge up process. The theoretical
$^{26}$Al/$^{27}$Al yield can be compared directly with observed yields in
stardust grains in meteoritic samples. For O-rich atmospheres considered in
this study, oxide grains need to be considered and from \citet{Nollett03} the
expected yield of $^{26}$Al/$^{27}$Al  in a typical low-mass 1.5M$\odot$ AGB star is seen to
lie between 10$^{-3}$ to 10$^{-1}$.
The highest values are reached after the star has undergone several thermal pulses -
a situation relevant for the evolved
OH/IR stars. In high mass AGB stars, the $^{26}$Al/$^{27}$Al yield may be even
higher \citep{Prantzos96}. At the low yield values of 10$^{-3}$, detecting the
the $^{26}$AlO component against the $^{27}$AlO emission background will be
challenging spectroscopically. However, at the higher yields the chance of a
detection should be good. In any case, even being able to observationally
set a well-constrained upper limit on the
$^{26}$Al/$^{27}$Al ratio should be useful (especially since the rate for the
key reaction $^{26}$Al(p,$\gamma$) $^{27}$Si that destroys $^{26}$Al
is largely uncertain; \citet{vanRaai08}). A robust $^{26}$Al detection should also help in
tightly constraining the  contribution of AGB stars to the galactic $^{26}$Al emission
manifested through the 1.8 MeV $\gamma$ ray line.

\acknowledgments
The research work at Physical Research Laboratory is funded by the Department
of Space, Government of India. Some of the data reported here were obtained
from time allocated from the UKIRT Service Programme for which we are
grateful. UKIRT is operated by the Joint Astronomy Centre on behalf of the
Science and Technology Facilities Council of the U.K. We are thankful to the
anonymous referee for helpful comments, graciously conveyed, that helped
improve the paper.


\begin{thebibliography}{}

\bibitem[Banerjee et al.(2003)]{Banerjee03} Banerjee, D. P. K., Varricatt,
  W. P., Ashok, N. M., Launila, O. 2003, \apj, 598, L31

\bibitem[Banerjee et al.(2004)]{Banerjee04} Banerjee, D. P. K., Ashok, N. M.,
Launila, O., Davis, C. J., Varricatt, W. P. 2004, \apj, 610, L29

\bibitem[Banerjee et al.(2005)]{Banerjee05} Banerjee, D. P. K., Barber, R. J.,
Ashok, N. M.,  Tennyson, J.,  2005, \apj, 627, L141

\bibitem[Banerjee et al.(2006)]{Banerjee06} Banerjee, D. P. K., Su, K. Y. L.,
Misselt, K. A., Ashok, N. M. 2006, \apj, 644, L57

\bibitem[Banerjee et al.(2007)]{Banerjee07} Banerjee, D. P. K., Misselt,
  K. A., Su, K. Y. L., Ashok, N. M., Smith, P. S. 2007, \apj, 666, L25

\bibitem[Benson et al.(1990)]{Benson90} Benson, P. J., Little-Marenin, I. R.,
Woods, T. C., Attridge, J. M. et al. 1990, \apjs, 74, 911

\bibitem[Bond et al.(2003)]{Bond03} Bond, H. E., Henden, A., Levay, Z. G.,
Panagia, N. et al. 2003, Nature, 422, 405

\bibitem[Bond et al.(2011)]{Bond11} Bond, H. E., 2011,  ApJ, 737, 17

\bibitem[Cernicharo et al.(1997)]{Cernicharo97} Cernicharo, J.,
Alcolea, J., Baudry, A., Gonzalez-Alfonso, E. 1997, \aap, 319, 607

\bibitem[Cesaroni et al.(1988)]{Cesaroni88} Cesaroni, R., Palagi, F.,
Felli, M., Catarzi, M., Comoretto, G., di Francos, Giovanardi, C., Palla, F. 1988, \aaps, 76, 445

\bibitem[Cheung et al.(1982)]{Cheung82} Cheung, A. S.-C., Taylor, A. W., Merer, A. J. 1982, JMoSp, 92, 391

\bibitem[Deguchi et al.(1989)]{Deguchi89} Deguchi, S., Nakada, Y., Forster,
  J. R. 1989, \mnras, 239, 825

\bibitem[Deguchi et al.(2000)]{Deguchi00} Deguchi, S., Fujii, T.,
Izumiura, H., Kameya, O., Nakada, Y., Nakashima, Jun-ichi., 2000, \apjs, 130, 351

\bibitem[Deguchi et al.(2002)]{Deguchi02} Deguchi, S., Fujii, T., Nakashima,
  Jun-Ichi., Wood, P. R. 2002, \pasj, 54, 719

\bibitem[Deguchi et al.(2004)]{Deguchi04} Deguchi, S., Fujii, T., Glass,
  I. S., Imai, H. et al. 2004, \pasj, 56, 765

\bibitem[Eder, Lewis \& Terzian(1988)]{Eder88} Eder, J., Lewis, B. M.,
  Terzian, Y., 1988, \apjs, 66, 183

\bibitem[Evans et al.(2003)]{Evans03} Evans, A., Geballe, T. R.,
Rushton, M. T., Smalley, B., van Loon, J. Th., Eyres, S. P. S.,
Tyne, V. H. 2003, \mnras, 343, 1054

\bibitem[Galehouse, Davis \&  Brault (1980)]{Galehouse80} Galehouse, D. C.,
  Davis, S. P., Brault, J. W. 1980, \apjs, 42, 241

\bibitem[Hansen \& Blanco(1975)]{Hansen75} Hansen, O. L., Blanco, V. M. 1975,
  \aj, 80, 1011

\bibitem[Hinkle et al.(1989)]{Hinkle89} Hinkle, K. H., Lambert, D. L., Wing, R. F. 1989, \mnras, 238, 1365

\bibitem[Humphreys et al.(2011)]{Humphreys11} Humphreys, R. M., Bond, H. E.,
  Bonanos, A. Z., Davidson, K., Berto Monard, L. A. G., Prieto, J. L., Walter, F. M., 2011, ApJ, 743, 118

\bibitem[Jorgensen(1994)]{Jorgensen94} Jorgensen, U. G. 1994, \aap, 284, 179

\bibitem[Joyce et al.(1998)]{Joyce98} Joyce, R. R., Hinkle, K. H., Wallace,
  L., Dulick, M., Lambert, D. L. 1998, \aj, 116, 2520

\bibitem[Keenan, Deutsch \& Garrison(1969)]{Keenan69} Keenan, P. C., Deutsch,
  A. J., Garrison, R. F. 1969, \apj, 158, 261

\bibitem[Kim et al.(2010)]{Kim10} Kim, J., Cho, Se-Hyung., Oh, C. S., Byun,
  Do-Young. 2010, \apjs, 188, 209

\bibitem[Launila \& Jonsson(1994)]{Launila94} Launila O., Jonsson, J. 1994, JMoSp, 168, 1

\bibitem[Launila \& Banerjee(2009)]{Launila09} Launila, O., Banerjee, D. P. K. 2009, \aap, 508, 1067

\bibitem[Le Bertre \& Nyman(1990)]{LeBertre90} Le Bertre, T., Nyman, L.-A. 1990, \aap, 233, 477

\bibitem[Le Squeren et al.(1992)]{LeSqueren92} Le Squeren, A. M., Sivagnanam,
  P., Dennefeld, M., David, P. 1992, \aap, 254, 133

\bibitem[Lewis, Eder \& Terzian(1990)]{Lewis90} Lewis, B. M., Eder, J.,
Terzian, Y. 1990, \apj, 362, 634

\bibitem[Luck \& Lambert(1974)]{Luck74} Luck, R. E., Lambert, D. L. 1974, \pasp, 86, 276

\bibitem[Lugaro \& Karakas(2008)]{Lugaro08} Lugaro, M., Karakas, A. I. 2008, NewAR, 52, 416

\bibitem[Lynch et al.(2004)]{Lynch04} Lynch, D. K., Rudy, R. J., Russell,
  R. W., Mazuk, S. et al. 2004, \apj, 607, 460

\bibitem[Munari et al.(2002)]{Munari02} Munari, U., Henden, A., Kiyota, S.,
Laney, D. et al. 2002, \aap, 389L, 51

\bibitem[Nollett, Busso \& Wasserburg(2003)]{Nollett03} Nollett,
  K. M., Busso, M., Wasserburg, G. J. 2003, \apj, 582, 1036

\bibitem[Nuth \& Hecht(1990)]{Nuth90}Nuth, J. A., III, \& Hecht, J. H. 1990, Ap\&SS, 163, 79

\bibitem[Nyman, Hall \& Olofsson(1998)]{Nyman98} Nyman, L.-A., Hall, P. J., Olofsson, H. 1998, \aaps, 127, 185

\bibitem[Olnon et al.(1986)]{Olnon86} Olnon, F. M. et al.  1986, A$\&$AS, 65, 607

\bibitem[Prantzos \& Diehl(1996)]{Prantzos96} Prantzos, N., Diehl, R. 1996, PhR, 267, 1

\bibitem[Siess \& Arnould(2008)]{Siess08} Siess, L., Arnould, M. 2008, \aap, 489, 395

\bibitem[Speck et al.(2000)]{Speck00} Speck, A. K., Barlow, M. J., Sylvester,
  R. J., Hofmeister, A. M. 2000, \aaps, 146, 437

\bibitem[Spencer et al.(1981)]{Spencer81} Spencer, J. H., Schwartz, P. R.,
Winnberg, A., Olnon, F. M., Matthews, H. E., Downes, D. 1981, \aj, 86, 392

\bibitem[te Lintel Hekkert et al.(1989)]{teLintelHekkert89} te Lintel Hekkert,
  P., Versteege-Hensel, H. A., Habing, H. J., Wiertz, M. 1989, A\&AS, 78, 399

\bibitem[te Lintel Hekkert et al.(1991)]{teLintelHekkert91} te Lintel Hekkert,
  P., Caswell, J. L., Habing, H. J., Haynes, R. F., Haynes, R. F., Norris, R. P. 1991, \aaps, 90, 327

\bibitem[Tenenbaum \& Ziurys(2009)]{Tenenbaum09} Tenenbaum, E. D., Ziurys,
  L. M. 2009, \apj, 694L, 59

\bibitem[van Raai et al.(2008)]{vanRaai08} van Raai, M. A., Lugaro, M.,
Karakas, A. I., Iliadis, C. 2008, \aap, 478, 521

\bibitem[Walker et al.(1997)]{Walker97} Walker, H. J., Tsikoudi, V., Clayton,
  C. A., Geballe, T., Wooden, D. H., Butner, H. M. 1997, \aap, 323, 442

\end{thebibliography}
\end{document}